\documentclass[12pt]{iopart}
 \usepackage{graphicx}
 \usepackage{dcolumn}
 \usepackage{bm}
\begin{document}
\title{Nuclear classical dynamics of H$_2$ in intense laser field} 

\author{Firoozeh Sami, Mohsen Vafaee, Babak Shokri}

\address{Laser-Plasma Research Institute, Shahid Beheshti University, G. C., Evin, Tehran 19839-63113, Iran}
\ead{ mo\_vafaee@sbu.ac.ir; Corresponding author}

\newcommand{\be}{\begin{equation}}
\newcommand{\ee}{\end{equation}}
\newcommand{\bea}{\begin{eqnarray}}
\newcommand{\eea}{\end{eqnarray}}
\newcommand{\h}{\hspace{0.30 cm}}
\newcommand{\vs}{\vspace{0.30 cm}}
\newcommand{\n}{\nonumber}

\renewcommand \thesection{\Roman{section}}
%----------------------------------------------------------------------------------
%%%%%%%%%%%%%%%%%%%%%%%%%%%%%%%%%%%%%%%%%%%%%%%%%%%%%%%%%%%%%%%
\begin{abstract}
In the first part of this paper, the different distinguishable pathways and regions of the single and sequential double ionization are determined and discussed.
It is shown that there are two distinguishable pathways for the single ionization and four distinct pathways for the sequential double ionization.
It is also shown that there are two and three different regions of space which are related to the single and double ionization respectively.
In the second part of the paper, the time dependent Schr\"{o}dinger and Newton equations are solved simultaneously for the electrons and the nuclei of H$_2$ respectively. 
The electrons and nuclei dynamics are separated on the base of the adiabatic approximation. 
The soft-core potential is used to model the electrostatic interaction between the electrons and the nuclei.
A variety of wavelengths (390 nm, 532 nm and 780 nm) and intensities ( $ 5\times10^{14}$ $Wcm^{-2} $ and $ 5\times10^{15}$ $Wcm^{-2} $) of the ultrashort intense laser pulses with a sinus second order envelope function are used.
%+++++++
%+++++++ the results of nuclear dynamics
%+++++++
The behaviour of the time dependent classical nuclear dynamics in the absence and present of the laser field are investigated and compared.
In the absence of the laser ﬁeld, there are three distinct sections for the nuclear dynamics on the electronic ground state energy curve.
%It is shown that there are three distinct sections on the electronic ground state energy curve for the nuclear dynamics in the absence of the laser field. 
The bond hardening phenomenon does not appear in this classical nuclear dynamics simulation.  
\\
\end{abstract}
%   \keywords{TDSE}
\maketitle
\section{Introduction}
Atoms and molecules exposed to intense laser pulses reveal a vast wealth of fascinating phenomena for example: single and double ionization \cite{Alnaser}, above-threshold ionization \cite{above-threshold ionization}, charge resonance enhanced ionization \cite{CREI}, dissociative-ionization \cite{Vafaee-8-10,Madsen_KER}, above threshold dissociation \cite{above-threshold dissociation}, bond softening and hardening \cite{Bandrauk1981, Frasinski1999, Magrakvelidze2009,Moser-Gibson} and high order harmonic generation \cite{HOHG}. Scientific research regarding these phenomena resulted in a revolution in ultrashort laser pulses and molecular science with many broad applications such as the control of the molecular processes \cite{control of molecular processes}, generation of a few cycles of femtosecond and attosecond pulses, the emergence of the attophysics \cite{attophysics}, and time-resolved imaging of molecular dynamics and reactions \cite{time-resolved imaging of molecular}.

Our knowledge about the interaction of atoms and molecules with ultra-short intense laser pulse has proceeded like essentially all other fields of atoms and molecules from simple systems such as Hydrogen atom and Hydrogen molecule to complex ones such as proteins. 
The first step in the chain of studying many-electron systems is investigation of the simplest two-electron systems, i.e. helium atom and H$_{2}$ diatomic molecule\cite{McKenna2009, Manschwetus2009, Dehghanian2010}.
The response of many electron atoms or molecules to the pulse of the laser field is often described by assuming that only one electron is active and responsible for the emission; this is called the single active electron (SAE) approximation. In this approximation, other electrons are assumed to contribute in the dynamics through a static screening potential. 
However, for full description and understanding the behaviour of the two-electron systems, it is necessary to consider both electron simultaneously without SAE approximation. This work plays a central role in developing our understanding of the interaction of many-electron atoms and molecules with ultrashort intense laser fields \cite{McKenna2009, Manschwetus2009, Dehghanian2010}. 
For this purpose, we need to solve the time-dependent Schr\"{o}dinger equation (TDSE) for the two-electron systems. This work is impossible with available current computing power.  
Therefore, at the present time unavoidably in many researches, modelling of the interaction of two-electron systems with the laser pulse is accomplished by the so-called soft-core Coulomb potentials which makes the numerical solution of TDSE possible \cite {Eberly1990, Kulander1996, Kstner2010, Camiolo, Saugout}.

In this work, we 
consider the nuclei as classical particles. This model reduces the complexity of the problem and helps to show the details of the dynamics of the electrons without the necessity that we get involved in complexities of the dissociative-ionization process \cite{Vafaee-8-10, Madsen_KER,Ergler2005, Staudte-Chelkowski2007, Litvinyuk2008, Jin-Tong2010}. In this research, the indistinguishability concept and the symmetry properties between two electrons will be demonstrated with some details. 
The main focus of attention in this paper is about the details of the classical nuclear dynamics and in a separate article the details of the electrons dynamics are represented \cite{VSSPRA}.

The paper is organized as follows: In the Sec. II and III, the details of the numerical solution of the TDSE and simulation box are described. In the Sec. IV, the results of the simulations for the field-free case and also for different intensities and wavelengths are presented and discussed. Finally, the conclusion appears in Sec.~IV.  
We use the atomic units ($\hbar = m_{e} = e = 1$) throughout this paper unless it is stated.
\section{Numerical solution of the TDSE}
\begin{figure}[h]
\begin{center}
	\resizebox{120mm}{!}{\includegraphics{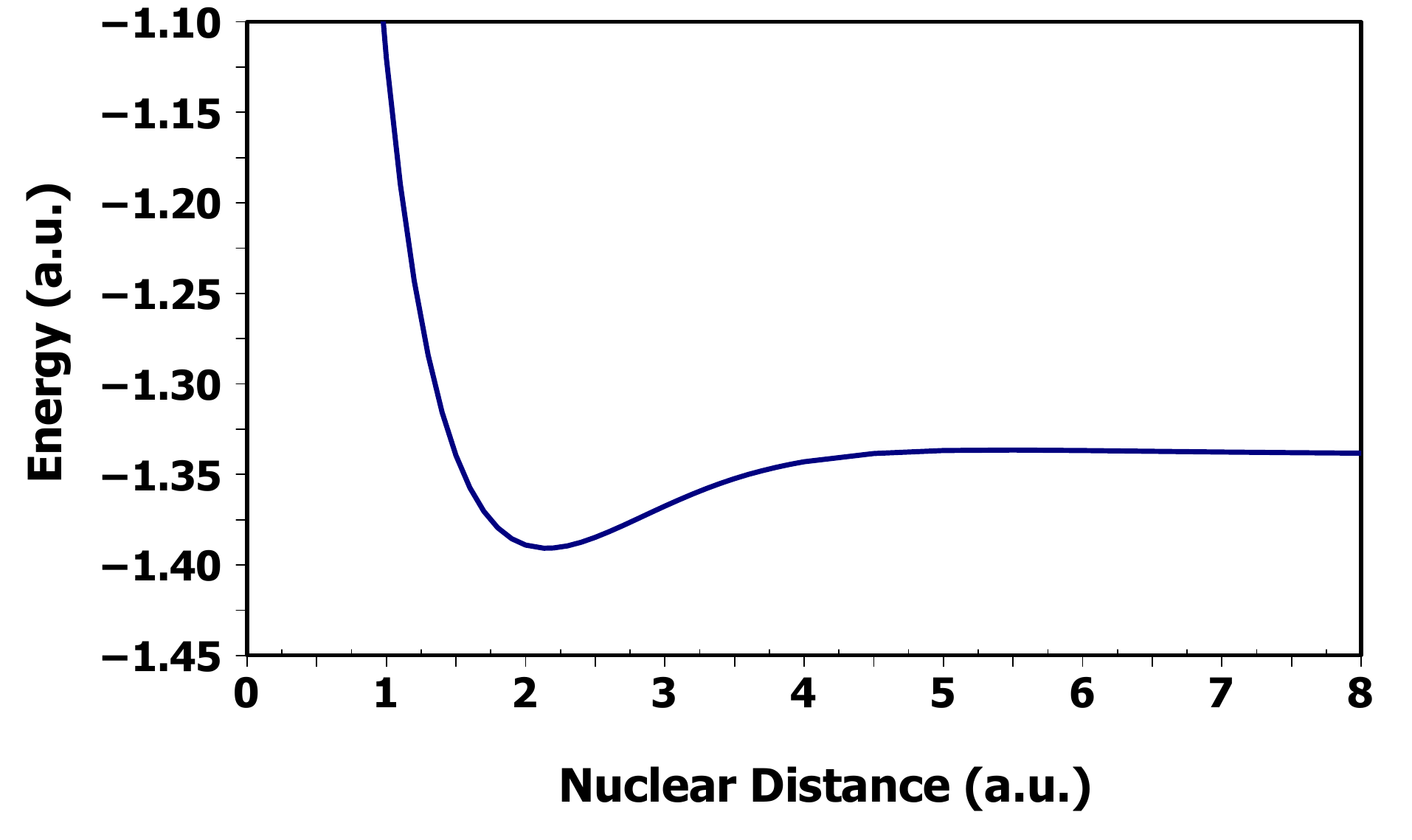}}
\end{center}
	  \caption
		{ 	\label{energypotential} 
Calculated the electronic ground states surface of hydrogen molecule on the base a one-dimensional model for both the electrons.
		}
\end{figure}

In this section we introduce the details of the numerical solution and also the simulation box that we used for investigating the dynamics of the hydrogen molecule in the presence of a strong laser field.
In a linearly polarized laser pulse with intensity upto $ 1\times10^{17}$ $Wcm^{-2} $, most of the electron and nuclei dynamics occur in the direction of the laser field \cite{Camiolo}. Therefore, we choose a one-dimensional model for both the electrons and nuclei coordinates. In what follows, $R_1$ and $R_2$ indicate the nuclei positions and $z_1$ and $z_2$ indicate the electron positions. Furthermore, $M$ and $m$ indicate the nuclei and electrons masses, respectively, and $–e$ is the electron charge. The temporal evolution of the electronic parts of such a system is described by the time dependent Schr\"{o}dinger equation (TDSE) on the base adiabatic approximation, i.e. \cite{Camiolo,Rigamonti}
\begin{eqnarray}\label{eq:1}
\lefteqn{i\frac{\partial \psi(z_1,z_2,t; R_1(t), R_2(t))}{\partial t}=}\nonumber\\
&&H_e(z_1,z_2,t; R_1(t), R_2(t)) \psi(z_1,z_2,t; R_1(t), R_2(t))
\end{eqnarray}
where the electronic Hamiltonian for this system, $H_e(z_1,z_2,t; R_1(t), R_2(t))$, is given by
\begin{eqnarray}
\lefteqn{H_e(z_1,z_2,t; R_1(t), R_2(t)) =}\nonumber\\
&&-\frac{1}{2m_{e}} \left[\frac{\partial ^2}{\partial z_1^2}+\frac{\partial ^2}{\partial z_2^2}\right]+V_{C}(z_1,z_2,t; R_1(t), R_2(t)),
%\nonumber\\&&
\label{eq:2}
\end{eqnarray}
\begin{eqnarray}
\lefteqn{V_{C}(z_1,z_2,t; R_1(t), R_2(t))=} \nonumber\\
&& \sum_{i, \alpha=1}^2 \left(\frac{-Z_\alpha}{\sqrt{( z_i-R_\alpha)^2+a }}\right)
 +\frac{1}{\sqrt{( z_1-z_2)^2+b }}\nonumber\\ 
 &&+\frac{Z_1Z_2}{\sqrt{( R_1-R_2)^2+c }}
+(z_1+z_2) E_0 f(t)\cos(wt),\nonumber\\
&&          
\label{eq:3}
\end{eqnarray}
where $ Z_1$ and $Z_2=1$ are the nuclei charge and $ a $, $ b $, and $ c $ are the soft-core parameters which are related to electron-nuclei, electron-electron, and nucleus-nucleus interaction, respectively . 
As this equatin shows, we use a soft-core potential to model the electrostatic interaction among the electrons and the nuclei by the screening parameters a, b and c. We choose $a=b=1.0$, $c=0.03$ \cite{Camiolo}.
The ground state energy and the equilibrium internuclear distance of $H_2$ become -1.39 and 2.13 respectively for this choice of the parameters (see Fig.~\ref{energypotential}).

\begin{figure}[ht]
\centerline{\includegraphics[width=10cm]{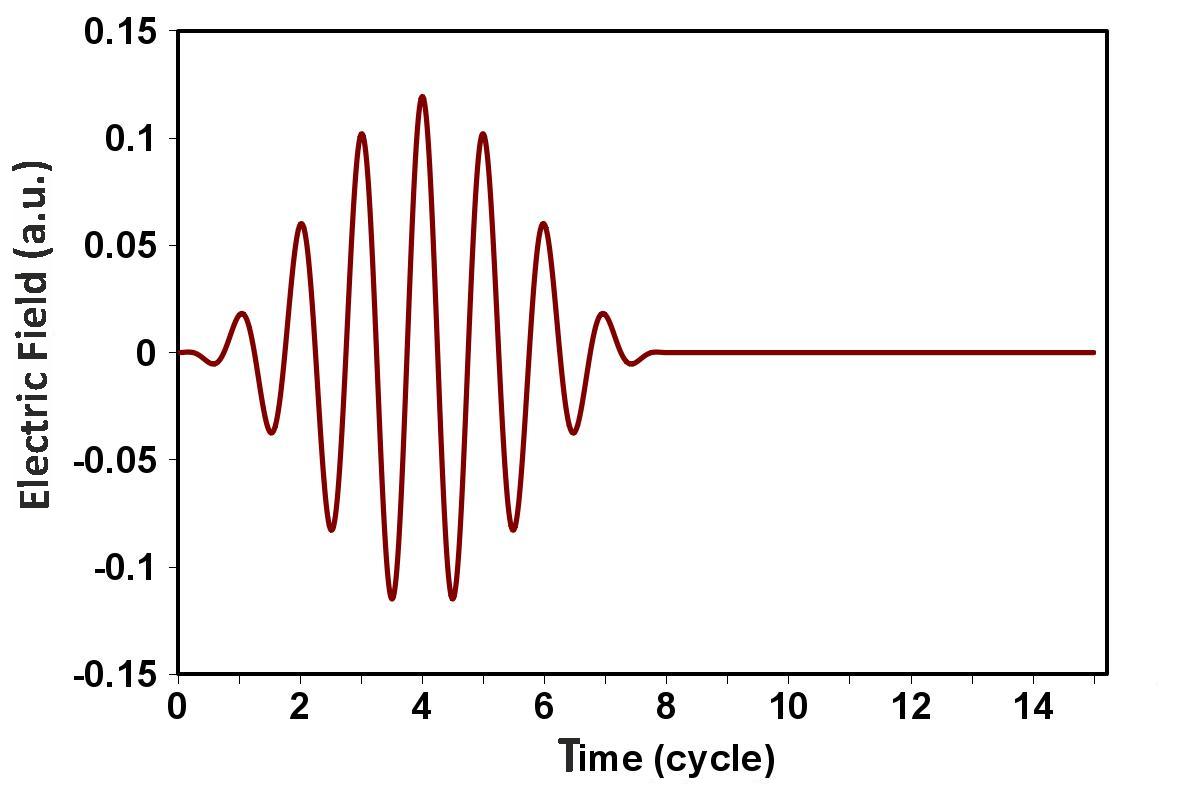}}
\caption{\label{field}\small The used laser electric field in this article has a sinus second order envelop function with 8 optical cycles. }
\end{figure}

The laser-molecule interaction is formulated in the dipole approximation in length gauge ($-e\textbf{r}.\textbf{E}(t)$), where $E_0$ is the laser peak amplitude and $\omega = 2\pi \nu $ is its angular frequency. The envelope of the laser pulse, $f(t)$, is set as
\begin{eqnarray}
f(t) =sin^{2}(\frac{t}{\tau_{1}}\pi)
\label{eq:4}
\end{eqnarray}
where $ \tau_1 $ is the time duration of the field irradiation, set at $ \tau_1 $=8 cycles in this work. 
1 cycle is equal to ~1.30 fs (~53.78 a.u.) and ~1.77 fs (~73.36 a.u.) for 390 nm and 532 nm respectively. 
After the mentioned time, the simulation continues for 8 more cycles in which there is no laser field as shown in Fig.~\ref{field} to follow the behavior of the system after turning off the laser field.
In this simulation, the time step is set to $ \delta t$= 0.02. The differential operators in Eq.~(2) are discretized by the 11-point difference formulas which have tenth-order accuracies \cite{vafa2006}. To solve the above TDSE numerically, we adopted a general nonlinear coordinate transformation for electronic coordinates. For the spatial discretization, we have constructed a finite difference scheme with a nonuniform (adaptive) grid for $ z_1$ and $z_2$ electronic coordinates, which are finest near the nuclei and coarsest at the border regions of the simulation box. More details of our calculations are described in our previous reports \cite{vafa2006,vafa2004}.
The absorber regions are introduced by using fourth-order optical potentials at the z$_1$ and z$_2$ boundaries, in order to capture the photoelectrons and prevent the reflection of the outgoing wave packets at the borders of the grid. More details of our calculations are described in our previous work \cite{vafa2004}. 

In this work, we separate the electrons and nuclei dynamics on the base of the adiabatic approximation \cite{Camiolo,Rigamonti}. We consider the electrons dynamics in the quantum approach while the nuclei dynamics are investigated in the classical manner. So, the time dependent Schr\"{o}dinger and Newton equations are solved simultaneously for the electrons and the nuclei respectively.
The equations govern the nuclear dynamics are as follows \cite{Camiolo}:
\begin{eqnarray}
MR_{1}(t)=F_{12}(t)+F_{1e}(t)+F_{L}(t)   \nonumber\\
MR_{2}(t)=F_{21}(t)+F_{2e}(t)+F_{L}(t)
\label{eq:5}
\end{eqnarray}
where  $ F_{L}(t)=e E_0 f(t)\cos(wt)$ is the laser forces exerted on each nucleus and $ F_{21}(t)=-F_{12}(t)$ are the internuclear repulsion 
\begin{eqnarray}
F_{21}(t)=-F_{12}(t)=\frac{R_{2}(t)-R_{1}(t)}{\left( c+\mid R_{2}(t)-R_{1}(t)\mid^{2}\right) ^{3/2}}
\label{eq:6}
\end{eqnarray}
and $ F_{ne} $ is the attractive forces between the electrons and nuclei where $ e $ and $ n $ are related to the electrons and nuclei respectively

\begin{eqnarray}
\lefteqn{F_{ne} (t)=-\int\int  \frac{[R_n (t)-z_1 ]\mid\Psi(z_1,z_2,t)\mid^2}{\left[ a+[R_n (t)-z_1 ]^2 \right] ^{3/2}}dz_1 dz_2}  \nonumber\\
&& -\int\int  \frac{[R_n (t)-z_2 ]\mid \Psi(z_1,z_2,t)\mid^2}{\left[ a+[R_n (t)-z_2 ]^2 \right] ^{3/2}}dz_1 dz_2.
\label{eq:6}
\end{eqnarray}
These Newtonian equations are solved by the Verlet algorithm \cite{Verlet}.  

\section{Different pathways and regions of the single and sequential double ionization}
\begin{figure*}[ht!]
\begin{center}
	\resizebox{120mm}{!}{\includegraphics{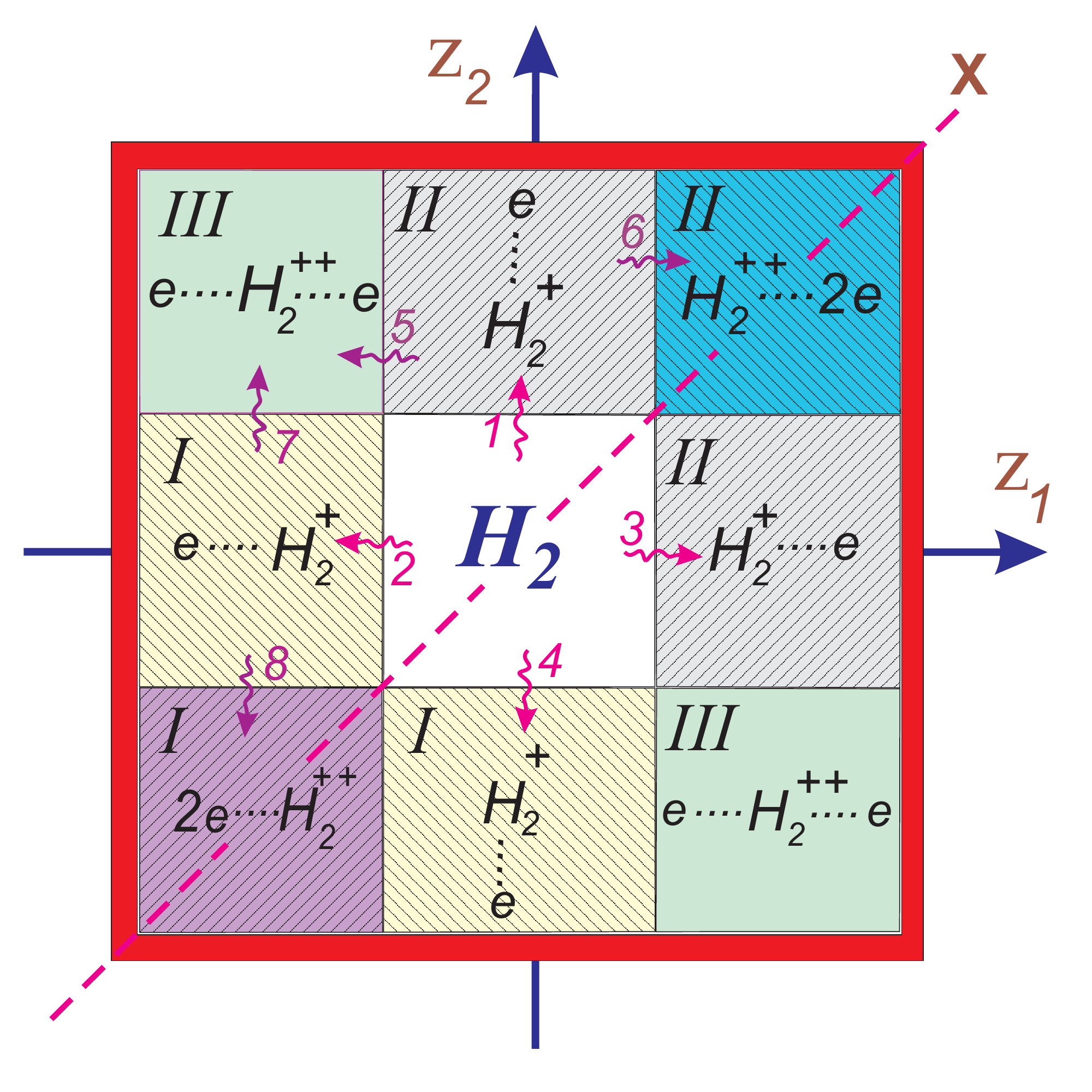}}
\end{center}
	  \caption
		{ 	\label{fig:f1} 
(Color online)  Schematic representation of the simulation box used in this work. It is constituted by two degrees of freedom pertinent to movement of the electrons along polarization direction.
One of electrons' direction adjoins with the horizontal axis ($ z_1 $) and the other is on the vertical one ($ z_2 $).
The simulation box are divided to some different regions related to the  H$_2$, H$_2^{+}$, and H$_2^{++}$.
The symmetry of wave-function in the box is identified by the diagonal line (X).
So, during computation, we can just consider the upside part of the simulation box with respect to the X line. 
According to indistinguishably, the behaviour of the downside of the box will be same as to the upside. 
The outer red regions surrounding the inner regions of the simulation box show the boundary absorption which lets absorption of the outgoing single and double ionization wavefunctions.
		}
\end{figure*}

In this section, we focus on different phenomena which might occur as a result of the interaction of the hydrogen molecule with the intense laser fields such as dissociation and serious kind of ionization process.
Due to the exclusion Pauli principle, the electronic wave function must always be antisymmetric. We assume an initial singlet two-electron state, so that the spatial part, $\psi(z_1,z_2,t; R_1(t), R_2(t))$, of the two electron wave function is symmetric.
As it is schematically shown in Fig.~\ref{fig:f1}, the simulation box is constituted by two degrees of freedom which are related to the movement of the electrons along polarization directions. One of electrons' direction adjoins with the horizontal axis ($ z_1 $) and the other is on the vertical one ($ z_2 $). Both of them include the negative and the positive halfs.  The symmetry of wave-function in the box is identified by the diagonal line (X) that traces from the down left corner to up right corner. This symmetry makes some opportunity for saving the necessary CPU time and memory in simulation and also increases the rate of the calculations. So, during computation, we can just consider the upside part of  the simulation box with respect to the X line in the Fig.~\ref{fig:f1}. According to indistinguishably, the behaviour of the downside of the box will be same as to the upside. 
For example, $\psi(z_1,z_2,t; R_1(t), R_2(t))$ of the H$_2^{+}$(II) regions in the upside part is equal to $\psi(z_2,z_1,t; R_1(t), R_2(t))$ of the H$_2^{+}$(II) for the downside part. 
Therefore, we calculate wave function for the upside part and derive the downside date from upside part.  
 
At the beginning time of the simulation, the hydrogen molecule is placed at an initial internuclear distance in the ground electronic state with the nuclei at rest. By passing the time, the density distribution of the electron clouds and also the position of the nuclei will be affected by the radiation of the laser field which leads to dissociation, ionization, and accomplishment phenomena. As long as the system is not ionized and dissociated, the population remains in the H$_2$ region of the Fig.~\ref{fig:f1}.

We now review some different kinds of outgoing of the electrons from region H$_2$. If due to the irradiation, just one of the two bounded electrons becomes affected; that electron goes away from the nuclei and the other one rests under the Coulomb attractive forces.
As a result of this process, the system goes out from the H$_2$ region and enters the neighbouring H$_2^+$ region. This process is called single ionization. 
 The single ionization can be occurred through four pathways which are shown in Fig.~\ref{fig:f1} by 1-4 arrows. Because of the indistinguishability, there are only two distinct pathways for the single ionization that can be presented by the arrows 1 and 2 or by the arrows 3 and 4. The pathways that are shown by arrows 3 and 4 are completely equivalent to arrows 1 and 2 respectively. In the pathway 1, the electron 2 ($e_2$) goes  away from nuclei in the positive direction while in the pathway 2 the electron 1 goes away in the negative direction which is in the opposite direction of the electron 2. 
The distinguishable movement in the pathway 1 and 2 is due to the symmetry breaking of the system by irradiation. 
Therefore, there are two distinguishable regions labelled I and II which are related to the H$_2^+$ region. 
In the absence of the laser field, the initial symmetric wavefunction remains symmetric with respect of the negative and positive half. 
Beyond the H$_2$ region, the force of the external field overcomes the Coulomb forces on the ionized electrons. Therefore, outside the H$_2$ region, the movement of the ionized electrons is controlled mainly by the laser field.
In the presence of the laser field after first ionization, it is expected that the next ionization take places. In this condition, the distance of the non-ionized electron increases and as a result the system goes out of the H$_2^+$(I) and H$_2^+$(II) regions which are related to the single ionization and enters the neighbouring H$_2^{++}$ regions. The pathways of the second ionization are represented in the upside region with respect to the X line and by arrows 5-8 in Fig.~\ref{fig:f1}. Therefore, there are four distinguishable pathways for the second ionization. This procedure leads to the sequential double ionization. 
Figure~\ref{fig:f1} shows three recognized different regions which are related to the double ionization, namely H$_2^{++}$(I), H$_2^{++}$(II), and H$_2^{++}$(III). 
The distinction of the H$_2^{++}$(I) and H$_2^{++}$(II) regions, like the H$_2^{+}$(I) and H$_2^{+}$(II) regions in Fig.~\ref{fig:f1}, is due to the radiation of the linearly polarized laser field that leads to the symmetry break between the left and right of the wavefunction. 
The behaviour of the system in the third region of the second ionization, i.e. the H$_2^{++}$(III) region, is completely different.
%The results of the simulation shows that the probability for the presence of the system in H$_2^{++}$(III) region is more than the H$_2^{+}$(I) and H$_2^{+}$(II) regions \cite{VSSPRA}.
Results in \cite{VSSPRA} (see Fig. 9) show a higher probability of finding the system in H$_2^{++}$(III) regions than in regions H$_2^{++}$(I) and H$_2^{++}$(II). 
This is due to the tendency of the system for remaining in a more stable region.% of the accessible state. 
When the system is in the H$_2^{++}$(III) region, the two electrons have gone away from the nuclei in the opposite directions and as a result the system has the least feasible inter-electronic repulsion. In the  H$_2^{++}$(I) and H$_2^{++}$(II) regions, the both electrons are in the same positive or negative direction. In this situation, the repulsive force between the two electrons makes the system more unstable. Therefore, it is expected that the double ionization with opposite directions (H$_2^{++}$(III)) is more probable than the double ionization with the same negative  (H$_2^{++}$(I)) or positive  (H$_2^{++}$(II)) directions. 
As shown in Fig.~\ref{fig:f1}, the boundary absorption has been placed in the end of the regions of the single and double ionization which lets absorption of the outgoing wavefunction.

The indistinguishability of two electrons remains unchanged in the intense laser field but the ionization process occurs via various distinguishable pathways.
There are two different pathways for the single ionization and  four different pathways for the double ionization. 
In the double ionization case, two electrons may go out from the right or left side of the simulation box. In the third pathway, the two electrons go out in the opposite sides. 

\section{Results and Discussion}

The electronic ground state of the hydrogen molecule is calculated by the imaginary time propagation method.
%The potential curves are presented in Fig.~\ref{energypotential}.
The obtained potential energy curve is presented in Fig.~\ref{energypotential}.
The electronic ground state energy is zero for the internuclear distance of the 0.45216 a.u..
The minimum energy is -1.39 a.u. and the related equilibrium distance for the classical nuclear motion is 2.13 a.u.. 

At first we study the field free behaviour of the system. 
The details of the size of the simulation box for the field free calculation is as follows.
The grid points for each z$_1$ and z$_2$ coordinates are 1110. The finest grid size value in an adaptive grid scheme is equal 0.2 a.u. for both z$_1$ and z$_2$ coordinates. The grids extend up to z$_1$ and z$_2$=$\pm114$ a.u.. The size of the absorber regions equals $\pm14$ a.u.. Therefore, the size of the simulation box equals $200\times200$, regardless of the absorber regions. In the present laser field, different size of the simulation boxes is used for the simulation box that is characterized.

\begin{figure}[h]
\begin{center}
	\resizebox{120mm}{!}{\includegraphics{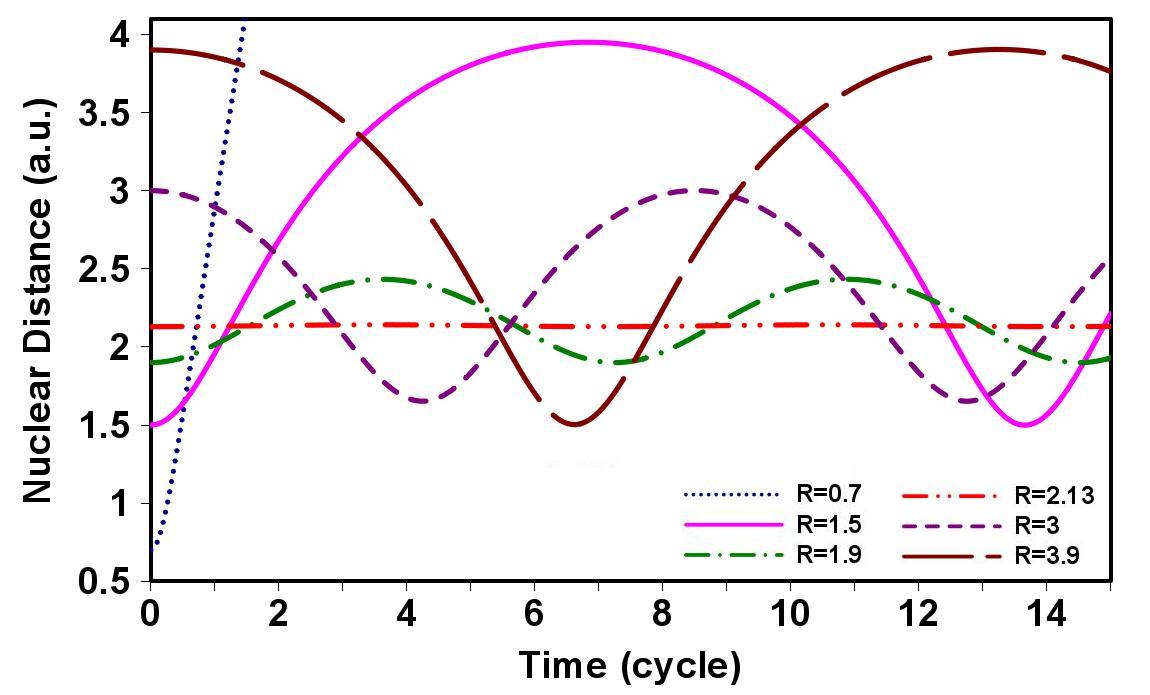}}
\end{center}
	  \caption
		{ 	\label{nuclearwf2} 
(Color online)  The time dependent internuclear distances belong to the second section (1.4 a.u.$<  $R$ _0<$5 a.u.) in a virtual laser field with a wavelength of 532 nm and zero intensity.
		}
\end{figure}

In the field free simulations, at the initial time, the system is released at the different internuclear distances, inner or outer turning points.
The calculation shows
three different sections on the electronic ground state energy curve. The first section is related to the internuclear distance smaller than 1.4 a.u. (R$ _0 <$1.4 a.u.), the next section is related to the interval between 1.4-5 a.u.(1.4 a.u.$<  $R$ _0<$5 a.u.) and the last section is for the internuclear distances which are greater than the 5 a.u. (R$ _0> $5 a.u.).
Since in the first section (R$_0< $1.4 a.u.), the nuclei are very close and the Coulomb repulsive force is so large, the system is unstable and the dissociation takes place easily. 
%The smaller the internuclear distance is, the dissociation process takes place faster. 
The smaller the internuclear distance, the faster dissociation process takes place.
However the reduction in the population of the electrons and the resultant ionization value is ignorable in this section. 
If the system is released in a internuclear distances belonging to the second section (1.4 a.u.$<  $R$ _0<$5 a.u.), which we call the potential well section, the ionization and dissociation do not take place which is shown in the Fig.~\ref{nuclearwf2}. By time evolution, the system just oscillates around the equilibrium internuclear distance. More close the initial internuclear distance to the equilibrium distance, the period of the oscillation decreases and as a result, the symmetry in the oscillation curve increases. At the equilibrium point, there is no oscillation.
The antisymmetric which is observed in the oscillation cycle in the initial internuclear distances which are far away from the equilibrium point, is due to the antisymmetric in the shape of the potential barrier walls. For example, the behavior of R$_{0}$=1.9 a.u. is relatively symmetric about the equilibrium but the behavior of 1.5 a.u. is completely antisymmetric about the equilibrium point (2.13 a.u.).
%The antisymmetric which is observed in the oscillation cycle in the internuclear distances which are far away from the equilibrium point, is due to the antisymmetric in the shape of the potential barrier walls.
Fig.~\ref{nuclearwf2} shows that the R$_0=$1.5 and 3.9 a.u. are the related inner and outer turning points. 
In the third section (R$>$5 a.u.) the Coulomb force is so weak and the bounding between the nuclei is not very strong,  so the dissociation does not take place unless the nuclei move to a greater internuclear distance at the initial time. 

\begin{figure*}[ht!]
\begin{center}
	\resizebox{150mm}{!}{\includegraphics{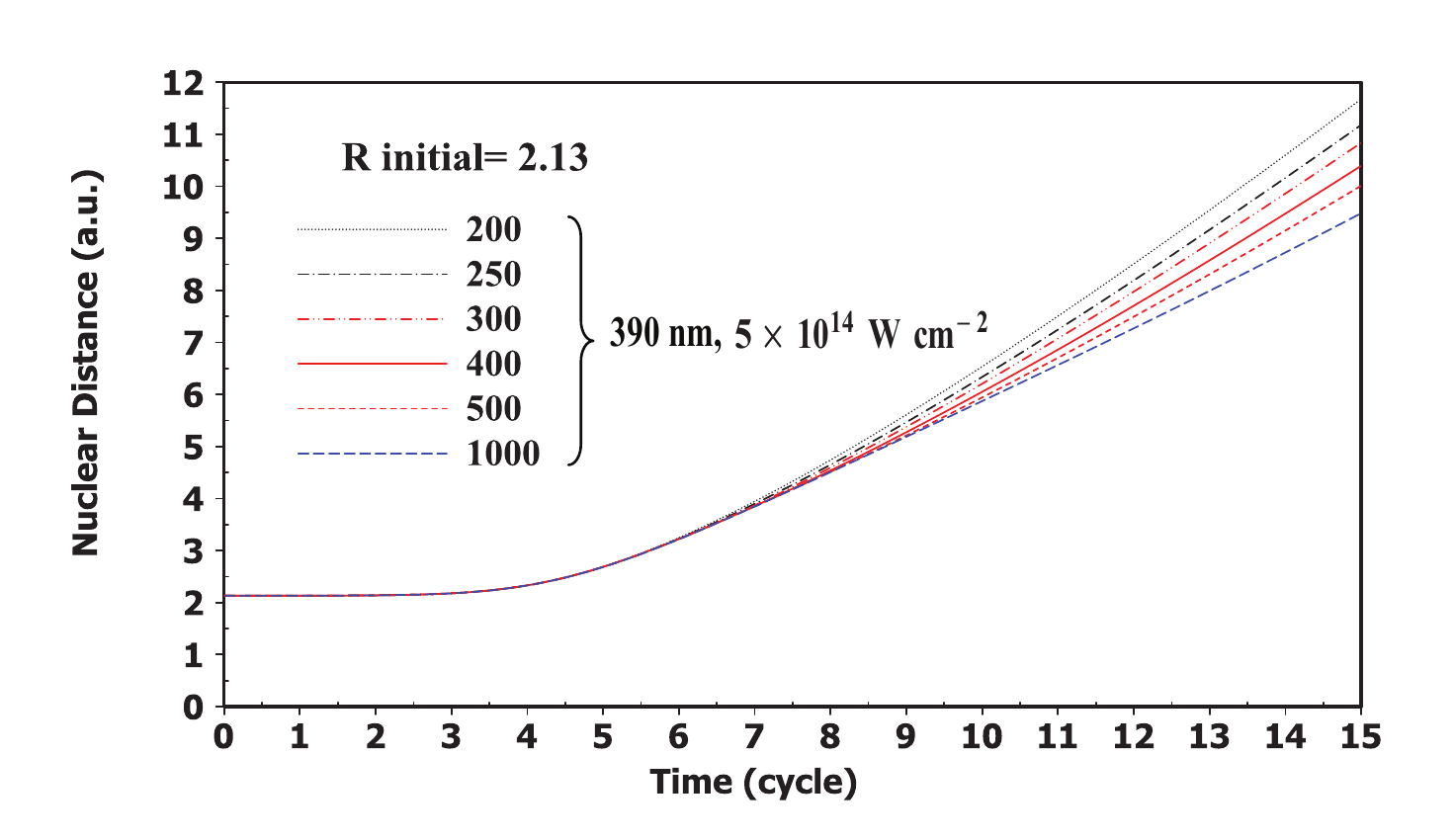}}
\end{center}
	  \caption
		{ 	\label{ND_conv} 
(Color online)  The time dependent internuclear distance for different sizes of the simulation box exposed to the laser field with a wavelength of 390 nm and $ 5\times10^{14}$ $Wcm^{-2} $ intensity. The results for the various sizes of the simulation box form $200\times200$ to $1000\times1000$ a.u.$^{2} $ are shown.  
		}
\end{figure*}

In the present of the laser field, at first we discuss the effect of the size of the simulation box on the nuclear dynamics. 
The calculations show that in the higher intensity ($ 5\times10^{15}$ $Wcm^{-2} $), the time dependent internuclear distance for different size of the simulation boxes is same as $200\times200$. 
In the lower intensity ($ 5\times10^{14}$ $Wcm^{-2} $), the results for the various sizes of the simulation box are shown in Fig.~\ref{ND_conv}. This figure shows that during 8 cycles of the laser pulse, there are a small difference in the internuclear distances for the various sizes. After 8 cycles, the difference in these internuclear distances become considerable and the increase in the size of the simulation box does not result in the convergence of the time dependent internuclear distance.
We changed our explanation in the article as below:
 “We can explain these results as follows. In the higher intensity of the laser pulse ($5\times10^{15}$ $Wcm^{-2}$), the increasing size of the simulation box does not effect on the nuclear dynamics. In this intensity, two electrons becomes completely far from nuclei during a few cycles of the laser pulse and a dominate Coulomb explosion occurs between nuclei. Therefore, the increasing size of the simulation box does not effect on the behavior of the nuclear dynamics. On the other side, at the lower intensity, the electrons do not move completely far from the nuclei. Therefore, both the Coulomb repulsion and the population of the electron about nuclei determine the time dependent internuclear distance. Therefore, when the size of the simulation box increases, the reminded population of the electrons about nucleus in simulation box before absorption by boundary absorption becomes slightly more and the magnitude of the Coulomb repulsion between nucleus is decreased.”

\begin{figure*}[ht!]
\begin{center}
	\resizebox{180mm}{!}{\includegraphics{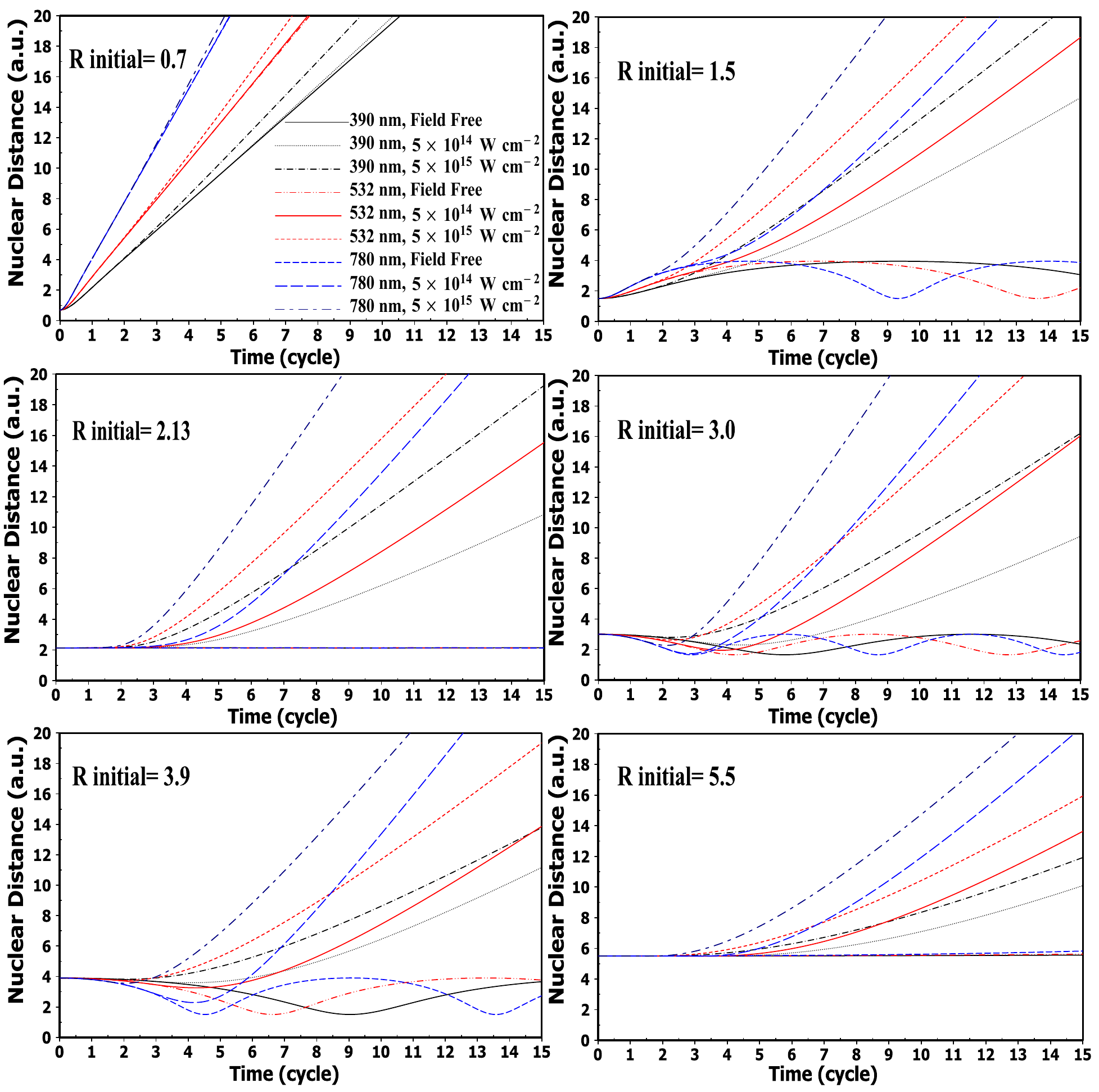}}
\end{center}
	  \caption
		{ 	\label{nudis} 
(Color online)  The time dependent internuclear distance for the three wavelengths and two different intensities and comparison with the results of the field-free in Fig.~ \ref{nuclearwf2}. 
		}
\end{figure*}

\begin{figure*}[ht!]
\begin{center}
	\resizebox{150mm}{!}{\includegraphics{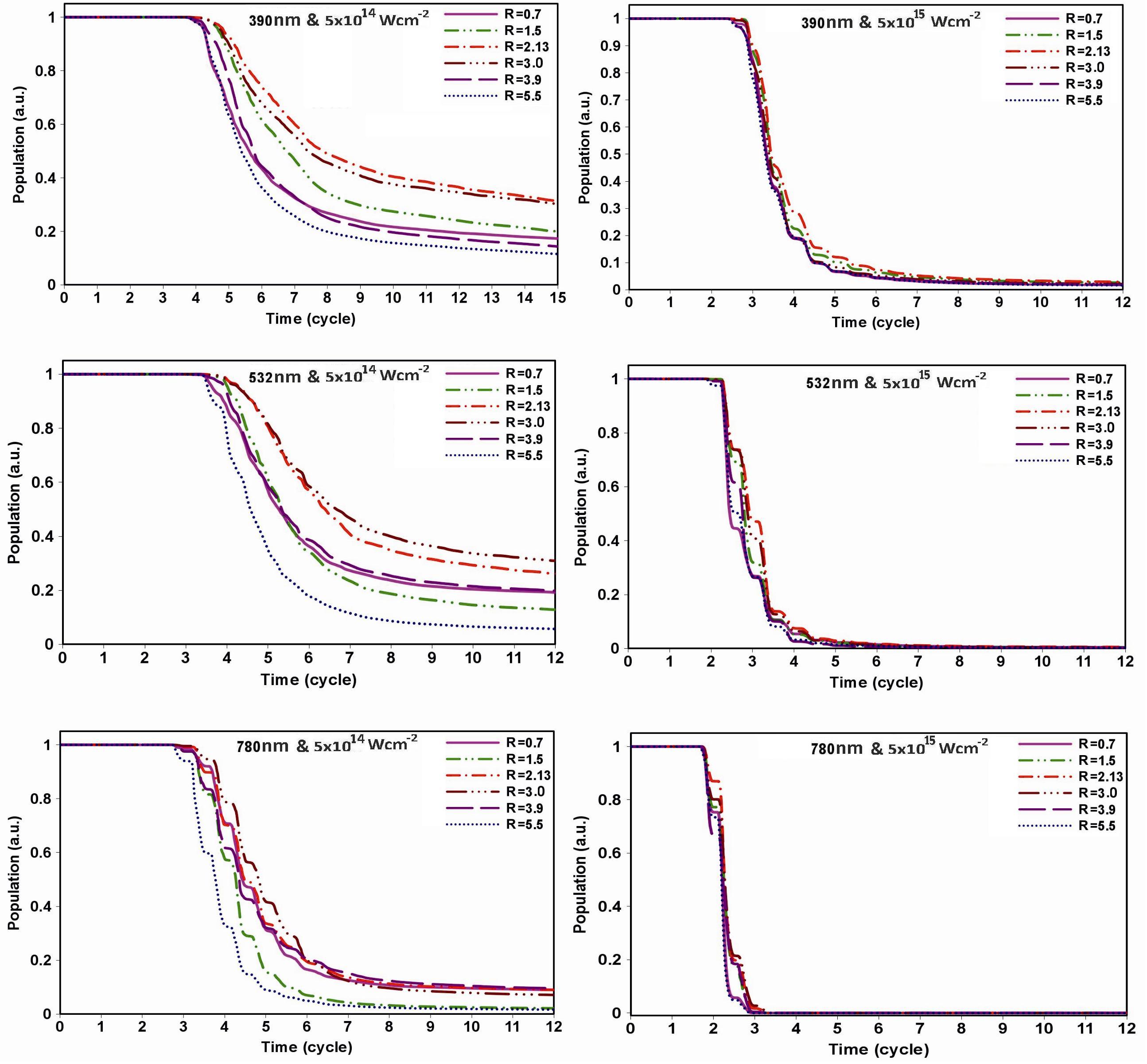}}
\end{center}
	  \caption
		{ 	\label{normcompar} 
(Color online)  The time dependent of the population for the different initial internuclear distance for the three wavelengths and two different intensities.
		}
\end{figure*}

The behaviour of the system under the action of the laser irradiation is different from what is seen in the absence of the laser field. 
Figure \ref{nudis} shows the time dependent internuclear distance for three different wavelengths and two different intensities. In this figure, the size of the simulation box for the $5\times10^{15}$ $Wcm^{-2}$ intensity is $200\times200$ and for the $5\times10^{14}$ $Wcm^{-2}$ intensity is $300\times300$.
%----
In addition, Fig.~\ref{normcompar} represents the time dependent behaviour of the population
(the residual norm in the simulation box that is equal to the total population of different regions in Fig.~\ref{fig:f1}) 
in the simulation box for three different wavelengths and the two intensities of the $ 5\times10^{14}$ $Wcm^{-2} $ and $ 5\times10^{15}$ $Wcm^{-2} $. In this figure, the size of the simulation box for both intensities is $200\times200$. 
The reduction of population is due to the outgoing of the electrons from the absorption boundaries, shown in Fig.~\ref{fig:f1}, that results to the single and double ionizations. In a separate article, we have shown the details of the single and double ionization \cite{VSSPRA}. 
The results of Fig.~\ref{nudis} and Fig.~\ref{normcompar} show respectively the magnitude of the internuclear distance (dissociation) and ionization for different sections and it is interesting to compare with the results of the field free case in Fig.~\ref{nuclearwf2}. 
%-----
In the first panel in Fig.~\ref{nudis} and ~\ref{normcompar}, i.e. R$ _0 =$0.7 a.u. that belongs to the first section, (R$ _0 <$1.4 a.u.), the dissociation occurs; just like what happens in the the absence of the laser field. We also observe a notable decline in the norm of the system which means that the electrons' population has gone out of the computation box and the ionization has occured. 
The panels with R$ _0 =$ 1.5, 2.13, 3.0, and 3.9 a.u. in  Fig.~\ref{nudis} and ~\ref{normcompar} belong to the second section, i.e. (1.4 a.u.$<  $R$ _0<$5 a.u.), and in contrast to the field free case both the ionization and dissociation have taken place. 

An interesting point in Fig.~\ref{nudis} is that the bond hardening phenomenon does not appear anywhere in these classical nuclear dynamics simulations. 
The bond hardening can take places when the laser field causes the nuclei to become closer to each other with respect of the field free case. 
The meaning of the bond hardening in the quantum nuclear dynamics investigation that was reported in the previous experimental and theoretical researches\cite{Bandrauk1981, Frasinski1999, Magrakvelidze2009} is vibrational trapping. In the vibrational trapping, the molecular wave packet is trapped in a laser-induced potential well. In contrary to the intuitive expectations, increasing the laser intensity can lead to the temporary stabilization of the molecular bond.
However, clear confirmation of bond hardening (or vibrational trapping) has remained elusive and might benefit from a fresh look \cite{Moser-Gibson}.

In the last panel with R$ _0 =$ 5.0 in Fig.~\ref{nudis} and ~\ref{normcompar} that is related to the third section (R$ _0> $5 a.u.), in contrast to the field free situation, both dissociation and ionization take place. In this section, the nuclei go away from each other slower in comparison with the second section, but the ionization rate is higher than the second section.

In all panels of  Fig.~\ref{nudis} and ~\ref{normcompar}, we can see that the electrons under the influence of the laser field go away from the nuclei which leads to the Coulomb explosion between the nuclei. 
The Coulomb repulsion for the second section is stronger than the third section which is due to the smaller initial internuclear distance in the second section when Coulomb expulsion is started.  

Figure~\ref{nudis} shows that the speed of the dissociation increase with increasing the wavelength or intensity of the laser field. 
This figure also shows that in the first section the wavelength is more effective than the intensity. In fact for the longer wavelengths, the nuclei separate from each other more quickly since in this case, the system is irradiated for the longer time in each half cycle of the laser field without a change in the direction of the laser field.
However, in the second section, the influence of the intensity overcomes the wavelength influence relatively. In fact in the second section, more the internuclear distance is close to the equilibrium point, 
the intensity increases the speed of the nuclei separation more.

Figure~\ref{normcompar} shows that the population reduction starts sooner with increasing the wavelength and the intensity. It should be mentioned that due to the increment of the stability of the system by getting closer to the initial equilibrium internuclear distance (R$_0=$2.13 a.u.), the population reduction is decreased.  
This figure also represents that in the second section for the smaller wavelengths, the ionization appears before beginning of the dissociation which shows that the laser irradiation makes Coulomb explosion by forcing the electrons' population to get away from the nuclei.   

%~~~~~~~~~~~~~~~~~~~~~~~~~~~~~~~~~~~~~~~
%^^^^^^^^^^^^^^^^^^^^^^^^^^^^^^^^^^^^^^

%
%
\section{Summary}
In this work and the following article \cite{VSSPRA}, we tried to fundamentally pay attention to this subject that how two electrons behave and interact in a two-electron molecule with the simplest classical manner of nuclear dynamics.  
In one part of this article, the distinguishable pathways and regions of the single and sequential double ionizations were determined and discussed. 
It is shown that the single ionization can occur through four pathways, but there are just two distinct pathways for the single ionization because of the indistinguishability of the electrons. 
This distinguishablity of the two pathways is due to the symmetry break between the right and left coordinates which is caused by the linearly polarized laser field. 
In the absence of the laser field, there is not such distinguishable pathways for electrons.
The result leads to the appearance of two distinguishable regions for the H$_2^+$ region in a linearly polarized intense laser pulse.
It is also shown that there are four distinct pathways for the ionization of the second electron.  
We showed that there are two different regions which are related to the single ionization and
three different ones that are related to the double ionization.
Among the regions of the second ionization, the H$_2^{++}$(III) region is more stable than the two other regions.
In this region, the two electrons go away from the nuclei in the opposite directions and as a result the system has the least feasible inter-electronic repulsion. However in the  H$_2^{++}$(I) and H$_2^{++}$(II) regions, the two electrons go away in the same positive or negative direction.
%+++++++
%+++++++ the results of nuclear dynamics
%+++++++

In the second part of this article, the time dependent Schr\"{o}dinger and Newton equations are solved simultaneously for the electrons and the nuclei of H$_2$ respectively. 
These calculations show that there are
three different sections on the electronic ground state energy curve in the absence of the laser field. 
In the first section (R$_0< $1.4 a.u.), the system is unstable and the dissociation takes place easily. However, the ionization value is ignorable in this section. 
In the potential well section (1.4 a.u.$<  $R$ _0<$5 a.u.) the ionization and dissociation do not take place. 
The system just oscillates around the equilibrium internuclear distance. More close the initial internuclear distance to the equilibrium distance, the period of the oscillation decreases and as a result, the symmetry in the oscillation curve increases. At the equilibrium point, there is no oscillation. 
In the third section (R$>$5 a.u.), the dissociation does not take place unless the nuclei move to a greater internuclear distance at the initial time.
In the present of the laser field, in all three section, the reduction of the electrons population is appeared in contrast to the field free case. 
The speed of the dissociation increase with increasing the wavelength or intensity of the laser field. 
In the first section the wavelength is more effective than the intensity.
In the second section, the influence of the intensity overcomes the wavelength influence.
In the third section (R$ _0> $5 a.u.), the nuclei go away from each other slower in comparison with the second section, but the ionization rate is higher than the second section.
In the present of the laser field, the time dependent nuclear distance is independent of the size of the simulation box at the higher intensity ($5\times10^{15}$ $Wcm^{-2}$) but at the lower intensity ($5\times10^{14}$ $Wcm^{-2}$), the final magnitude of the nuclear distance depends on the size of the simulation box. 
Finally, the bond hardening phenomenon does not appear anywhere in these classical nuclear dynamics simulations.  

\ack
%\begin{acknowledgments}
%
We thank Dr. Niknam for using his computing facilities and Ms. Mehrabian for her comments and careful reading of the manuscript. We wish to acknowledge Shahid Beheshti University for the financial supports, research facilities and High Performance Computing cluster of Laser and Plasma Research Institute(HPCLP).
%
%\end{acknowledgments}
%
%==================================================================================================


\Bibliography{<num>}
%\bibliography{p7}% Produces the bibliography via BibTeX.
% \begin{thebibliography}{100}

\bibitem{Alnaser}
 A. S. Alnaser, T. Osipov, E. P. Benis, A. Wech, C. L. Cocke, X. M. Tong and C. D. Lin, Phys. Rev. Lett. {\bf 91}, 163002 (2003).
% A. S. Alnaser et al, Phys. Rev. Lett. {\bf 93}, 183202 (2004);

\bibitem{above-threshold ionization} 
T. Brabec (Ed.), Strong Field Laser Physics , Springer Series in Optical Sciences (Springer, 2008).

\bibitem{CREI}  
T. Zuo and A. D. Bandrauk, Phys. Rev. A {\bf 52}, R2511 (1995);
G. N. Gibson, M. Li, C. Guo, and J. Neira, Phys. Rev. Lett. {\bf 79},
2022 (1997).
\bibitem{Vafaee-8-10}
M. Vafaee, Phys. Rev. A {\bf78}, 023410 (2008);
M. Vafaee, B. Shokri, Phys. Rev. A {\bf81}, 053408  (2010)
\bibitem{Madsen_KER}
H. A. Leth, L. B. Madsen, and K. Mølmer, Phys. Rev. Lett. {\bf103}, 183601 (2009);
H. A. Leth, L. B. Madsen, and K. Mølmer, Phys. Rev. A {\bf81}, 053409 (2010);
H. A. Leth, L. B. Madsen, and K. Mølmer, Phys. Rev. A {\bf81}, 053410 (2010).
%Dissociative double ionization of H2 and D2: Comparison between experiment and Monte Carlo wave packet calculations
%^^^^^^^^^^^^^^^^^^^^^^^^^^^^^^^^^^^^^^^^

\bibitem{above-threshold dissociation} 
A. Giusti-Suzor, F. H. Mies, L. F. DiMauro, E. Charron and B. Yang, J. Phys. B {\bf 28}, 309 (1995);
J. H. Posthumus, Rep. Prog. Phys. {\bf 67} 623 (2004);
B. D. Esry and I. Ben-Itzhak, Phys. Rev. A {\bf 82}, 043409 (2010);
Zheng-Tang Liu, Kai-Jun Yuan, Chuan-Cun Shu, Wen-Hui Hu and Shu-Lin Cong, J. Phys. B {\bf 43}, 055601 (2010).

%\bibitem{bond hardening}
\bibitem{Bandrauk1981}A. D. Bandrauk and M. L. Sink, J. Chem. Phys. {\bf 74}, 1110 (1981).
\bibitem{Frasinski1999}L. J. Frasinski, J. H. Posthumus, J. Plumridge, K. Codling, P. F. Taday, and A. J. Langley, Phys. Rev. Lett. {\bf 83}, 3625 (1999),
\bibitem{Magrakvelidze2009}M. Magrakvelidze, F. He, T. Niederhausen, I. V. Litvinyuk, and U. Thumm, Phys. Rev. A {\bf 79}, 033410 (2009).
\bibitem{Moser-Gibson}
B. Moser and G. N. Gibson, Phys. Rev. A {\bf 80}, 041402 (2009). 
%Ultraslow dissociation of the H2+ molecular ion via two-color ultrafast laser pulses

\bibitem{HOHG} 
P B Corkum, Phys. Rev. Lett. {\bf 71}, 1994 (1993);
M. Lewenstein , P. Balcou , M. Y. Ivanov, A. L’Huillier and P. B. Corkum, Phys. Rev. A {\bf 49}, 2117 (1994).

%\bibitem{review} 
%F. Grossmann, Theoretical Femtosecond Physics: Atoms and Molecules in Strong Laser Fields (Springer-Verlag, 2008);
%M. Lein, J. Phys. B {\bf 40}, R135 (2007); 
%M. Yu. Ivanov, A. Scrinzi, R. Kienberger and D. M. Villeneuve, J. Phys. B {\bf 39}, R1 (2006).

\bibitem{control of molecular processes}
M. F. Kling et al., Science {\bf 312}, 246 (2006).
\bibitem{attophysics}
T. Brabec and F. Krausz, Rev. Mod. Phys. {\bf 72}, 545 (2000);
T. Kanai, S. Minemoto, and H. Sakai, Nature (London) {\bf 435}, 470 (2005);
F. Krausz and M. Ivanov, Rev. Mod. Phys. {\bf 81}, 163 (2009).
%Attosecond physics
\bibitem{time-resolved imaging of molecular}
C. Z. Bisgaard et al., Science {\bf 323}, 1464 (2009);
S. Haessler, J. Caillat, W. Boutu, C. Giovanetti-Teixeira, T. Ruchon, T. Auguste, Z. Diveki,
P. Breger, A. Maquet, B. Carr\'{e} R. Ta\"{i}eb, and P. Sali\'{e}res, Nature Phy. {\bf 6}, 200 (2010).
% Attosecond imaging of molecular electronic wavepackets 

%\bibitem{two-elctron}
\bibitem{McKenna2009} J. McKenna, A. M. Sayler, B. Gaire, Nora G. Johnson, K. D. Carnes, B. D. Esry, and I. Ben-Itzhak
Phys. Rev. Lett. {\bf 103}, 103004 (2009);
%Benchmark Measurements of H3+ Nonlinear Dynamics in Intense Ultrashort Laser Pulses

\bibitem{Manschwetus2009} B. Manschwetus, T. Nubbemeyer, K. Gorling, G. Steinmeyer, U. Eichmann, H. Rottke, and W. Sandner, Phys. Rev. Lett. {\bf 102}, 113002 (2009);
%Strong Laser Field Fragmentation of H2: Coulomb Explosion without Double Ionization

\bibitem{Dehghanian2010} E. Dehghanian and A. D. Bandrauk, and G. Lagmago Kamta, Phys. Rev. Lett. {\bf 81} 061403(R) (2010).
%Enhanced ionization of the H2 molecule driven by intense ultrashort laser pulses

%\bibitem{soft-core}
\bibitem{Eberly1990} J. H. Eberly, Phys. Rev. A {\bf 42}, 5750 (1990).
\bibitem{Kulander1996} K. C. Kulander, F. H. Mies, and K. J. Sch\''{a}fer, Phys. Rev. A {\bf53},
2562 (1996).
\bibitem{Kstner2010} A. K\"{s}tner, F. Grossmann, R. Schmidt, and J. M. Rost, Phys. Rev. A {\bf 81}, 023414 (2010).
%Reliability of soft-core approximations in theoretical studies of molecules in intense laser fields

\bibitem{Camiolo}
 G. Camiolo, G. Castiglia, P. P. Corso, E. Fiordilino, and J. P. Marangos, Phys. Rev. A {\bf 79}, 063401 (2009).

\bibitem{Saugout}
 S. Saugout, E. Charron, and C. Cornaggia, Phys. Rev. A {\bf 77}, 023404 (2008).
%---------------KER-----------
\bibitem{Ergler2005}
T. Ergler, A. Rudenko, B. Feuerstein, K. Zrost, C. D. Schr\"{o}ter, R. Moshammer, and J. Ullrich, Phys. Rev. Lett. {\bf95}, 093001 (2005).
\bibitem{Staudte-Chelkowski2007}
 A. Staudte, D. Pavicic, S. Chelkowski, D. Zeidler, M. Meckel, H. Niikura, M. Sch\"{o}ffler, S. Sch\"{o}ssler, B. Ulrich, P. P. Rajeev, Th. Weber, T. Jahnke, D. M. Villeneuve, A. D. Bandrauk, C. L. Cocke, P. B. Corkum, and R. D\"{o}rner, Phys. Rev. Lett. {\bf 98}, 073003 (2007).
 %A. Staudte et al., Phys. Rev. Lett. {\bf98}, 073003 (2007);
S. Chelkowski, A. D. Bandrauk, A. Staudte, and P. B. Corkum,
Phys. Rev. A {\bf76}, 013405 (2007).
\bibitem{Litvinyuk2008}
I. V. Litvinyuk and A. S. Alnaser and D. Comtois and D. Ray and A. T. Hasan and J-C kieffer and D. M. Villeneuve, New J. Phys. {\bf 10}, 083011 (2008).
\bibitem{Jin-Tong2010}
Y. J. Jin, X. M. Tong, and N. Toshima, Phys. Rev. A {\bf81}, 013408 (2010) .
%Enhanced ionization of hydrogen molecular ions in an intense laser field via a multiphoton resonance
\bibitem{VSSPRA}  M. Vafaee, F. Sami and B. Shokri, arXiv:1012.4063v1.

\bibitem{Rigamonti}
A. Rigamonti and P. Carretta, Structure of Matter: An Introductory Course with Problems and Solutions  (Springer-Verlag, 2009).

\bibitem{vafa2006}  M. Vafaee, H. Sabzyan, Z. Vafaee and A. Katanforoush, arXiv:physics/0509072v4.
\bibitem{vafa2004}  M. Vafaee and H. Sabzyan, J. Phys. B {\bf 37}, 4143 (2004).

\bibitem{Verlet}
L. Verlet, Phys. Rev. {\bf 159}, 98 (1967); Phys. Rev. {\bf 165}, 201 (1967). 


%\end{thebibliography}
\endbib
\end{document}